\documentclass[12pt]{iopart}
\usepackage[T1]{fontenc}
\usepackage[latin9]{inputenc}
\usepackage[english]{babel}
\usepackage{graphicx}
\usepackage{setspace}
\usepackage{babel}
\begin{document}

% Make ket |psi>
\global\long\def\psiket{\left|\psi\right>}
% Make bra <psi|
\global\long\def\psibra{\left<\psi\right|}
% Make ket for spin up along axis #1
\global\long\def\upket#1{\left|\uparrow_{#1}\right>}
% Make ket for spin down along axis #1
\global\long\def\downket#1{\left|\downarrow_{#1}\right>}
% Make bras for up and down
\global\long\def\upbra#1{\left<\uparrow_{#1}\right|}
\global\long\def\downbra#1{\left<\downarrow_{#1}\right|}
% Make inner product <#1|#2>
\global\long\def\innerproduct#1#2{\left<#1|#2\right>}
% Define style for operator - in case journal wants particular format
% for operators, so I don't have to change them one by one
\global\long\def\myop#1{#1}
% Define style for vector - again in case journal wants particular format
\global\long\def\myvec#1{\vec{#1}}
% Define 2 element column vector
\global\long\def\colvec#1#2{\left(\begin{array}{c} #1 \\ #2 \end{array}\right)}
% Define 4x4 matrix
\global\long\def\mat#1#2#3#4{\left(\begin{array}{cc} #1 & #2 \\ #3 & #4\end{array}\right)}

\title{Applying Classical Geometry Intuition to Quantum Spin}

\author{Dallin S.\ Durfee and James L.\ Archibald\footnote{Currently at AMS-TAOS USA Inc.}}
%\email{dallin\_durfee@byu.edu}
\address{Department of Physics and Astronomy, Brigham Young University, Provo,
Utah 84602}
\ead{dallin\_durfee@byu.edu}

\begin{abstract}
Using concepts of geometric orthogonality and 
linear independence, we
logically deduce the form of the Pauli spin matrices and the
relationships between the three spatially orthogonal
basis sets of the spin-1/2 system. Rather than a mathematically rigorous 
derivation, the relationships are found by forcing expectation values of
the different basis states to have the properties we expect
of a classical, geometric coordinate system.
The process highlights the correspondence
of quantum angular momentum with classical notions of geometric orthogonality,
even for the inherently non-classical spin-1/2 system.  In the process,
differences in and connections between geometrical space
and Hilbert space are illustrated.
\end{abstract}

Keywords: spin, quantum mechanics, orthogonality

\maketitle

\section{Introduction}

It is possible to find the Pauli spin matrices
and the corresponding geometrically
orthogonal basis sets for the spin-1/2 system
using arguments from classical geometry.
Like the vector model \cite{beiser,LohSpinCoherentState},
this undertaking leverages classical understanding
to build intuition for quantum angular momentum.
But rather than using a semi-classical model, 
our approach uses completely quantum spin states.
It involves the application of classical principles to expectation values,
in a manner similar to Ehrenfest's theorem \cite{Ehrenfest,ThompsonAngularMomentum}.
The technique can provide valuable insight for
anyone with an advanced undergraduate or introductory
graduate understanding of quantum mechanics.

This method illuminates the correspondence between
Hilbert space, which is used to mathematically describe 
quantum states, and the geometrical space we use in
classical descriptions of angular momentum.
It makes connections between the parameters used
to define a Cartesian coordinate system
and the free parameters selected when defining geometrically
orthogonal sets of basis states to describe
a quantum spin-1/2 particle.  
The procedure clarifies some
common misconceptions about the quantum picture of angular 
momentum, in particular the confusion students often have
distinguishing between geometric and Hilbert space,
discussed in \cite{SternGerlach}.
It does not explain the \emph{why} of quantum spin, 
but focuses on how we
\emph{work with} quantum angular momentum states.

There are several rigorous ways to derive the well-known relationships 
between the $x$, $y$, and $z$ basis sets of the spin-1/2 system and
the corresponding Pauli matrices.  
These include the use of
rotation operations \cite{MerzbacherRotation,SakauraiRotation},
ladder operators \cite{GriffithsLadderOperator}, direct diagonalization \cite{NarducciDirectDiagonalization},
or symmetry arguments \cite{DiracBookSpinOfElectron,EpsteinSpinWOCommutation}.
Here we 
\emph{deduce}, rather than \emph{derive}, the relationships.

We limit our discussion
to a spin-1/2 system for simplicity and clarity.  This also avoids 
dealing with different types of unbiased states \cite{AngleStates}.
For example, as discussed later, for a spin-1/2 particle described using the
$z$ basis, the only states with a zero expectation value of the $z$ component
of angular momentum are ones made of an equal superposition of spin up and spin down.
For a spin-1 particle, this is not only
achieved with an equal superposition of $m=-1$ and $m=1$, but also
with the $m=0$ basis state, or
any combination of all states for which the -1 and 1 basis states have amplitudes of
equal size.
 
The canonical spin-1/2 system
illustrates the universality of the core principle behind Ehrenfest's theorem, even
in systems with no classical analog  \cite{OhanianWhatIsSpin,SpinHistory};
even though you cannot 
explain an electron's intrinsic angular momentum as
the result of physical rotation in the 
classical sense, the expectation
values of a spin-1/2 system do obey principles of classical geometry.
While the {\em origin} of
spin cannot be described semi-classically,
classical intuition {\em can}
be applied to better understand the {\em consequences} of spin.

\section{Quantum Spin-1/2 Formalism}

It is common to describe spin-1/2 particles using a basis consisting
of the two eigenstates of the $z$-component of angular momentum,
$\upket z$ and $\downket z$.  
In vector notation, we can write these states as
\begin{equation}
\upket z = \colvec 1 0 \qquad \mbox{and} \qquad \downket z = \colvec 0 1 .
\end{equation}
We will use the $z$ basis as
our starting point to deduce the form of the Pauli spin matrices
and find the relationships between the $x$, $y$, and $z$ basis
states.  

We assume that the $z$ basis exists and
that every possible
state of our particle can be expressed as a sum of these two states
in the form 
\begin{equation}
\psiket=a\upket z+b\downket z = \colvec a b ,\label{eq:Psi}
\end{equation}
where $a$ and $b$ are scalar, potentially complex, constants.

In matrix notation, the operators that yield information about the $x$,
$y$, and $z$ components of spin can be written as
$\hbar/2$ times the Pauli spin matrices.  The Pauli spin 
matrix $\sigma_z$ can be easily found by noting that
it is the matrix for which
$\upket z$ and $\downket z$ are the eigenvectors with eigenvalues
of +1 and -1.  This gives us the matrix
\begin{equation}
{\sigma_z = \mat 1 0 0 {-1} .}
\end{equation}

If we measure the $z$-component of angular momentum for
a spin-1/2 particle in an arbitrary state, we will always get either plus
or minus $\hbar/2$, with probabilities of $a^{*}a$ and $b^{*}b$
if the state is normalized.
To predict the outcome of a measurement along a \emph{different} axis,
we can write our quantum state in terms of the
eigenstates of the component of angular momentum
we plan to measure.
Since the $z$ axis is chosen arbitrarily, 
we know that
a pair of eigenstates of \emph{any} component of angular momentum 
along \emph{any} axis must exist 
and have similar properties to the $z$ basis states.

Two special basis sets are made up of the eigenstates of angular momentum
along the $x$ and $y$ axes, respectively. Because the $z$ basis
forms a complete set, we should be able to write these basis states
in terms of the $z$ states \cite{note:xandycomplete}. 
We'll start by writing the $x$ basis states as
\begin{eqnarray}
\upket x & = & A\upket z+B\downket z\mbox{ and }\label{eq:xupguess}\\
\downket x & = & C\upket z+D\downket z\label{eq:xdownguess}
\end{eqnarray}
where $A$, $B$, $C$, and $D$ are constants. 
We can deduce what these constants must be simply by
considering what
properties these states should have.
In the process, we will
see connections between physical (geometric) space and Hilbert space.

\section{Orthogonality}

To find $A$ and $B$ in Eq.\ \ref{eq:xupguess} we note that
a classical
particle with its angular momentum in the $x$ direction
will have no component of angular momentum in the $z$ direction.
For a quantum spin-1/2 particle we 
will \emph{always} measure the $z$ component
to be $\pm \hbar/2$, never zero, regardless
of the particle's state. So rather than mapping our intuition
of geometrical orthogonality onto possible measurement outcomes, 
we'll
make our $x$ basis states geometrically orthogonal
to $z$ by setting the \emph{expectation value} of the $z$ component
of angular momentum to zero.

For the expectation value to be zero, 
it must be exactly as likely to measure the $z$ component of spin
to be $-\hbar/2$ as it is to measure $+\hbar/2$, such that
the two possibilities cancel each other out.  As such,
we intuitively expect that $\upket x$ should be an equal
superposition of $\upket z$ and $\downket z$.

To show this
more rigorously, we use the operator
\begin{equation}
\myop{S_{z}}=\frac{\hbar}{2}\left(\upket z\upbra z-\downket z\downbra z\right) = \frac{\hbar}{2} \sigma_z
\end{equation}
to find the expectation value of the $z$ component of angular momentum.
Using this operator,
we get 
\begin{equation}
\upbra x\myop{S_{z}}\upket x=\frac{\hbar}{2}\left(A^{*}A-B^{*}B\right).\label{eq:Zcomponent}
\end{equation}
We require that this expectation value be zero, such that 
\begin{equation}
A^{*}A-B^{*}B=|A|^{2}-|B|^{2}=0.
\end{equation}
As we expected, $A$ and $B$ must have the same magnitude, giving
an equal superposition of $\upket z$ and $\downket z$. But the
complex phases of $A$ and $B$ are unrestricted. So the most general
form for $\upket x$, subject only to the limitation that
it be normalized and geometrically orthogonal to the $z$ basis states,
is 
\begin{equation}
\upket x=\frac{1}{\sqrt{2}}\left[e^{i\phi_{1}}\upket z+e^{i\phi_{2}}\downket z\right]
\end{equation}
where $\phi_{1}$ and $\phi_{2}$ are arbitrary real constants.

\begin{figure}
\begin{centering}
\includegraphics[width=5cm]{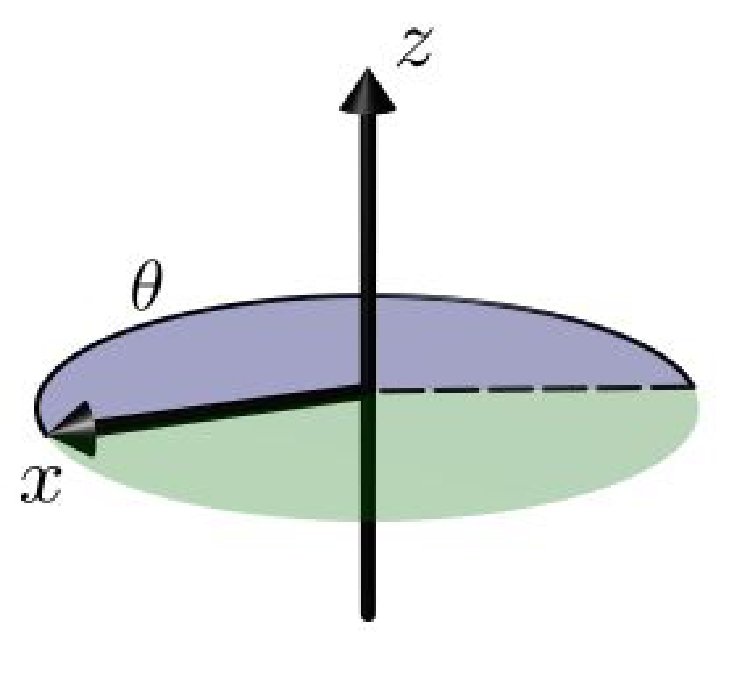}
%width=8.4 cm fills column
\par\end{centering}
\caption{Once the $z$ axis of a Cartesian coordinate system is selected, there
are still an infinite number of possible choices for the $x$ axis
which are all orthogonal to $z$. These choices are parametrized by
the variable $\theta$ in the figure. (Color online)\label{fig:definingx}}
\end{figure}

It makes sense that we get one arbitrary phase angle, since quantum
mechanics always allows us an arbitrary overall phase factor. But
why two? There's a geometric explanation for this freedom. After defining
the $z$ axis of a Cartesian coordinate system, the $x$ axis can
point in any one of an infinite number of directions which are orthogonal
to $z$. These choices can be parametrized by an angle relative to
some reference direction, as shown in Fig.\ \ref{fig:definingx}.

We will set $\phi_{1}$ and $\phi_{2}$ 
to zero - both for simplicity and because it
gives us the conventional form of the basis state,

\begin{equation}
{\upket x=\frac{1}{\sqrt{2}}\left[\upket z+\downket z\right].\label{eq:xup}}
\end{equation}

\section{Linear Independence}

To find $\downket x$ we note that it, too, must be geometrically
orthogonal to the $z$ basis. So it must have the form

\begin{equation}
\downket x=\frac{1}{\sqrt{2}}\left[e^{i\phi_{3}}\upket z+e^{i\phi_{4}}\downket z\right].
\end{equation}
We find an additional constraint when we consider that the $x$ basis
states \emph{could} have been the $z$ basis states had we simply
chosen a different direction for our $z$ axis. As such, since our
$z$ basis states are orthogonal to each other, we expect the two
states in the $x$ basis to be orthogonal to each other as well. 

As is often done, we've unfortunately used the word
``orthogonal'' to mean two different things. When we say that
the $x$ states must be orthogonal to the $z$ states, we are referring
to geometric orthogonality. But when we say that the $x$ states must
be orthogonal to each other, we refer to \emph{orthogonality in Hilbert
space} or 
\emph{linear independence}.  Just as a dot product of zero
assures geometric orthogonality, an inner product
$\innerproduct{\uparrow_{x}}{\downarrow_{x}}$ = 0 guarantees 
that $\upket x$ and $\downket x$ are 
linearly independent, such that one can't be written in terms
of the other.

The inner product of the $x$ basis states is
\begin{eqnarray}
\innerproduct{\uparrow_{x}}{\downarrow_{x}} & = &
\frac{1}{2}\left[\vphantom{e^{i\phi_{3}}}\upbra z+\downbra z\right]
\left[e^{i\phi_{3}}\upket z+e^{i\phi_{4}}\downket z\right] \nonumber \\
& = & \frac{1}{2}\left(e^{i\phi_{3}}+e^{i\phi_{4}}\right).
\end{eqnarray}
For this to be zero, the two phases must differ by $\pi$ radians.
If we choose the arbitrary global quantum phase of this state such that
$\phi_{3}=0$, both for simplicity and by convention,
we find that $e^{i\phi_{4}}=-1$ and
\begin{equation}
{
\downket x=\frac{1}{\sqrt{2}}\left[\upket z-\downket z\right].\label{eq:xdown}
}
\end{equation}

Knowing the form of $\upket x$ and $\downket x$, 
it is simple to show that the Pauli spin matrix 
$\sigma_x$ is given by
\begin{equation}
{\sigma_x = \mat 0 1 1 0 .}
\end{equation}

\section{The $y$ basis}

Just as we saw for the $x$ basis states, for the $y$ basis states to be normalized and geometrically
orthogonal to the $z$ axis, they must have the form
\begin{eqnarray}
\upket y & = & \frac{1}{\sqrt{2}}\left[e^{i\phi_{5}}\upket z+e^{i\phi_{6}}\downket z\right]\label{eq:yupguess}\\
\downket y & = & \frac{1}{\sqrt{2}}\left[e^{i\phi_{7}}\upket z+e^{i\phi_{8}}\downket z\right].\label{eq:ydownguess}
\end{eqnarray}
Since we have the freedom to multiply each state 
by an arbitrary overall phase
factor, for simplicity (and to arrive at the canonical form of
the states), we can set $\phi_{5}$ and $\phi_{7}$ to zero.
The other phase angles are constrained by that fact
that, in addition to being geometrically orthogonal to $z$, these
states must also be geometrically orthogonal to the $x$ axis we've
defined. 

The operator which gives us information about the $x$ component
of spin can be found by noting that, when written in the $x$ basis,
this operator should look similar to $\myop{S_{z}}$ represented in
the $z$ basis:
\begin{equation}
\myop{S_{x}}=\frac{\hbar}{2}\left(\upket x\upbra x-\downket x\downbra x\right).
\end{equation}
To write this in the $z$ basis we plug in Eqs.\ (\ref{eq:xup})
and (\ref{eq:xdown}) to get
\begin{equation}
\myop{S_{x}}=\frac{\hbar}{2}\left(\upket z\downbra z+\downket z\upbra z\right) = \frac{\hbar}{2}\sigma_x.
\end{equation}

We can use $\myop{S_{x}}$ to find the expectation value of the $x$
component of angular momentum for a particle in the state $\upket y$:
\begin{equation}
\upbra y\myop{S_{x}}\upket y  =  \frac{\hbar}{4}\left(e^{i\phi_{6}}+e^{-i\phi_{6}}\right) =  \frac{\hbar}{2}\cos\left(\phi_{6}\right).\label{eq:sx}
\end{equation}
If we want $\upket y$ to be geometrically orthogonal to
the $x$ states then this must be zero, implying that $\phi_{6}=\pm\pi/2$
and giving us only two unique possibilities: 
\begin{equation}
\upket y=\frac{1}{\sqrt{2}}\left[\upket z\pm i\downket z\right].\label{eq:yuppm}
\end{equation}
As we discuss in the next section, the choice of whether to use the
upper or lower sign is not arbitrary, so we'll not select one over
the other just yet. 

Applying the same condition to $\downket y$ and
forcing the inner product $\innerproduct{\uparrow_{y}}{\downarrow_{y}}$
to be zero gives us
\begin{equation}
\downket y=\frac{1}{\sqrt{2}}\left[\upket z\mp i\downket z\right].\label{eq:ydownpm}
\end{equation}

\section{Coordinate system handedness}

\begin{figure}
\begin{centering}
\includegraphics[width=4cm]{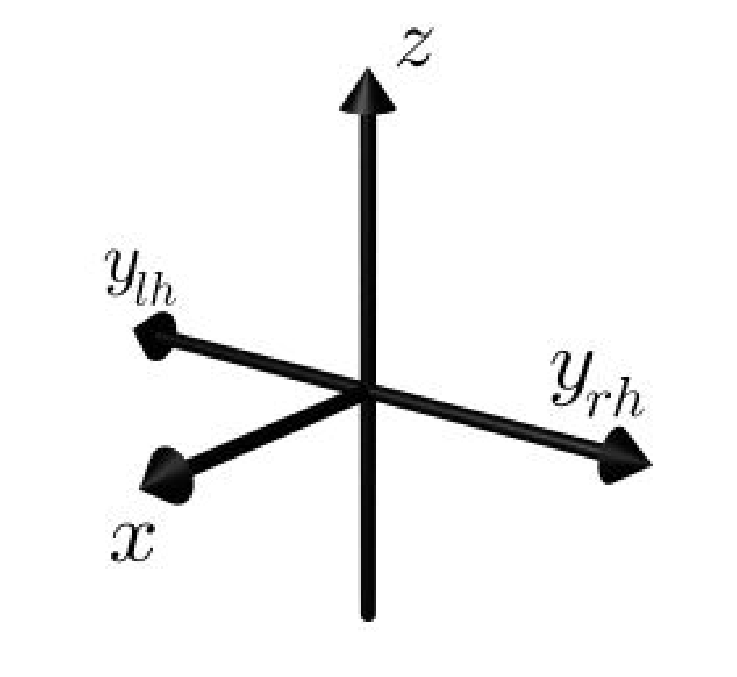}
%width=8.4 cm fills column
\par\end{centering}

\caption{Once the $x$ and $z$ axes are selected, there are still two possible
choices for the $y$ direction. One choice, labeled $y_{lh}$ in the
figure, will result in a left-handed coordinate system. The other
choice results in a right-handed coordinate system. \label{fig:definingy}}
\end{figure}

There is a geometric explanation for the two possible choices in Eqs.\
(\ref{eq:yuppm}) and (\ref{eq:ydownpm}). In a Cartesian coordinate
system, once the $z$ and $x$ axes have been selected, there are
still two possible directions for $y$. As illustrated in Fig.\ \ref{fig:definingy},
one direction results in a right-handed and the other in a left-handed
coordinate system. 

We can find the handedness of a geometric coordinate system
with cross products.  
For example, if $\hat{n}_{j}$ is a unit vector along the $j^{th}$ axis, then
for a right-handed coordinate system using a right-handed cross product, 
$\hat{n}_x \times \hat{n}_y = \hat{n}_z$. 
For a left-handed coordinate system we get a minus sign:
$\hat{n}_x \times \hat{n}_y = - \hat{n}_z$.

Similarly, we can find the ``handedness''
of the Pauli spin matrices by 
noting that the Pauli matrices, multiplied
by a constant and combined with
the ($2 \times 2$) identity matrix, form the basis
of the SU(2) Lie group \cite{su2}.
Then we can evaluate Lie brackets, 
which are analogous to cross products.
If we evaluate the Lie bracket of $\sigma_z/2$ with $\sigma_x/2$, 
we get
\begin{equation}
\left[\frac{1}{2} \sigma_z, \frac{1}{2} \sigma_x \right] = \frac{i}{2}\sigma_{y+} = - \frac{i}{2}\sigma_{y-} \label{eq:liebracket}
\end{equation}
where $\sigma_{y+}$/$\sigma_{y-}$ is the Pauli spin matrix we get if we choose the upper/lower
sign for Eqs.\ \ref{eq:yuppm} and \ref{eq:ydownpm}.  The
change in sign in Eq.\ \ref{eq:liebracket}
is just what we would expect when going from a right-
to a left-handed coordinate system.

Similarly, there is a link between the Pauli
matrices and quaternions \cite{QuaternionPauli, Gough86, Penrose97}, and between cross
products and commutators in quaternion algebra \cite{QuaternionCalculus}.
This suggests a connection between cross products
in geometric space and commutators in Hilbert space.

Because there is no spatial representation of the
spin-1/2 particle's angular momentum operator, we'll make an
analogy with the orbital angular momentum operator  \cite{DiracBookAngularMomentum}:
\begin{equation}
\myvec{\myop L}=\myvec{\myop r}\times\myvec{\myop p}.
\end{equation}
Here $\vec{r}$ is the position and $\vec{p}$ the momentum operator.
Assuming a right-handed coordinate system, the components of $\myvec L$
are
\begin{eqnarray}
\myop{L_{x}} & =\myop y\myop{p_{z}}-\myop z\myop{p_{y}}\\
\myop{L_{y}} & =\myop z\myop{p_{x}}-\myop x\myop{p_{z}}\\
\myop{L_{z}} & =\myop x\myop{p_{y}}-\myop y\myop{p_{x}}.
\end{eqnarray}

We can use these components to calculate
the commutator of $\myop{L_{x}}$ with $\myop{L_{y}}$.
Noting that position and linear momentum operators commute with operators
for orthogonal spatial dimensions and that $[\myop z,\myop{p_{z}}]=i\hbar$,
it is easy to show that $ \left[\myop{L_{x}},\myop{L_{y}}\right] =  i\hbar\myop{L_{z}}$.
If we had chosen a left-handed coordinate system (but still used
a right-handed cross product), we would have gotten the same
result but with a minus sign.

By analogy, we may suppose that a right-handed coordinate system for
a spin-1/2 particle is the one that results in the commutation relation
\begin{equation}
[\myop{S_{x}},\myop{S_{y}}]=i\hbar\myop{S_{z}}, \label{eq:RHScommutator}
\end{equation}
while a left-handed coordinate system would result in a minus sign
in the commutation relation.
We get the commutation relation in Eq.\ \ref{eq:RHScommutator}
if we choose the upper sign in Eqs.\ (\ref{eq:yuppm})
and (\ref{eq:ydownpm}):
\begin{eqnarray}
\upket y=\frac{1}{\sqrt{2}}\left[\upket z+ i\downket z\right]\\
\downket y=\frac{1}{\sqrt{2}}\left[\upket z- i\downket z\right].
\end{eqnarray}
From these, the Pauli matrix $\sigma_y$ can be found:
\begin{equation}
\sigma_y = \mat 0 {-i} i 0 .
\end{equation}
This is the last piece of the puzzle, 
and we have now ``deduced'' the relationships between the $x$, $y$,
and $z$ basis states as well as the Pauli matrices for the spin-1/2 system.

\section{Conclusion}

We deduced the Pauli matrices and the
relationships between the $x$, $y$, and $z$ basis states
for a spin-1/2 particle
using the concepts of geometric orthogonality and linear independence.
By insisting
that the two states in the $x$ basis be normalized, geometrically orthogonal
to the $z$ states, and orthogonal to each other in Hilbert space (linearly independent),
we arrived at expressions which were completely specified except for
three arbitrary phase angles - two due to arbitrary overall phase factors,
and a third related to choosing the direction for the $x$ axis
for a given selection of $z$ axis direction
in a Cartesian coordinate system.

With the $y$ basis we had less freedom because the states had
to be geometrically orthogonal to both the $z$ \emph{and} 
the $x$ basis states.
We again had
an arbitrary overall phase factor for each basis state. But we only
had two possible choices for the remaining phase factors,
similar to the choice of handedness in a Cartesian coordinate system.
We determined handedness by making a connection between
cross products and commutators.

This intuitive exercise illustrates the connection between
classical geometric space and quantum Hilbert space, even
for spin-1/2 systems, which are intrinsically not classical.

\section*{Acknowledgments}

We thankfully acknowledge
Jean-François Van Huele for very helpful discussions, and we are
grateful to Christopher
J. Erickson and Manuel Berrondo for feedback on this manuscript.
This research was funded by NSF Grant PHY-1205736 and by BYU's College
of Physical and Mathematical Sciences.

\section*{References}

%\bibliographystyle{phaip}
%\bibliographystyle{unsrt}
%\bibliographystyle{aipnum4-1}

%\bibliographystyle{iopart-num}
%\bibliography{SpinOneHalf_2016_03_28_EPL}

\begin{thebibliography}{10}
\bibitem{note:xandycomplete}
And, likewise, since the $x$ and $y$ basis sets will be complete, we should be
  able to write the $z$ basis states in terms of the $x$ or $y$ basis states.
\end{thebibliography}

\providecommand{\newblock}{}

\end{document}